\documentclass[referee]{aa} 
\usepackage{graphicx}
\usepackage{txfonts}
\usepackage{verbatim}
\usepackage{amssymb}
\usepackage{natbib}
\bibpunct{(}{)}{;}{a}{}{,}
\begin{document}
   \title{First results from a VLBA proper motion survey of H$_2$O masers in 
low-mass YSOs: the Serpens core and RNO~15-FIR.}

   \titlerunning{VLBA of water masers towards low-mass YSOs}

   \subtitle{}

   \author{L. Moscadelli\inst{1}
          \and
          L. Testi\inst{2}
          \and
          R.S. Furuya\inst{3}
          \and
          C. Goddi\inst{1}
	  \thanks{e-mail: cgoddi@ca.astro.it}
          \and
          M. Claussen\inst{4}
          \and
          Y. Kitamura\inst{5}
          \and
          A. Wootten\inst{4}
          }

   \offprints{L. Moscadelli}

   \institute{INAF, Osservatorio Astronomico di Cagliari,
              Loc. Poggio dei Pini, Str. 54, 09012 Capoterra (CA), Italy \\
              \email{mosca@ca.astro.it}
         \and
             INAF, Osservatorio Astrofisico di Arcetri, 
	     Largo E. Fermi 5, 50125 Firenze, Italy\\
             \email{lt@arcetri.astro.it}
         \and
             Division of Physics, Mathematics, and Astronomy,
             California Institute of Technology, MS 105-24,
             Pasadena, CA 91125, USA \\
             \email{rsf@astro.caltech.edu}
         \and
             National Radio Astronomy Observatory
             USA\\
         \and
             Institute of Space and Astronautical Science, Yoshinodai 3-1-1, Sagamihara,
             Kanagawa 229-8510, Japan
             }

   \date{}

   \abstract{
    This article reports first results of a long-term observational
    program aimed to study the earliest evolution of jet/disk systems
    in low-mass YSOs by means of VLBI observations of the 22.2~GHz
    water masers. We report here data for the cluster
    of low-mass YSOs in the Serpens molecular core and for the single
    object RNO~15-FIR. Towards Serpens SMM1, the most luminous sub-mm
    source of the Serpens cluster, the water maser emission comes from
    two small ($\leq$ 5~AU in size) clusters of features separated by
    $\approx$25~AU, having line of sight velocities strongly 
    red-shifted (by 
    more than 10~km~s$^{-1}$) with respect to the LSR velocity of the 
    molecular cloud. The two maser clusters are oriented on the sky
    along a direction that is approximately perpendicular to the axis
    of the radio continuum jet observed with the VLA towards SMM1. The spatial
    and velocity distribution of the maser features lead us to favor the 
    interpretation that the maser emission is excited by interaction 
    of the
    receding lobe of the jet with dense gas in the accretion
    disk surrounding the YSO in SMM1. The line of sight velocities of
    several features decrease at a rate of $\approx$1~km~s$^{-1}$~
    month$^{-1}$ and the sky-projected relative motion of two features  
    appears to be accelerated (decelerated)
    at a rate of $\approx$10--15~km~s$^{-1}$~month$^{-1}$.  
    We propose 
    that the shocks harboring the maser emission are slowed down as 
    they proceed through the dense material surrounding the YSO.  
    Towards RNO~15-FIR, the
    few detected maser features have both positions and (absolute) 
    velocities aligned along a direction that is parallel to the axis of the
    molecular outflow observed on much larger angular scales. In this case
    the maser emission likely emerges from dense, shocked molecular clumps 
    displaced along the axis of the jet emerging from the YSO.
    The protostar in Serpens SMM1 is more massive than the one in RNO~15-FIR.
    We discuss the case where a high mass ejection rate can generate
    jets sufficiently powerful to sweep away from their course
    the densest portions of circumstellar
    gas. In this case, the excitation
    conditions for water masers might preferably occur at the interface
    between the jet and the accretion disk, rather than along the jet axis.    
   }

   \maketitle
%

\section{Introduction}

   The scenario based on the disk/jet system for the formation of an isolated 
   low-mass star  \citep{Shu87} is supported both by millimeter interferometer
   (i.e. \citet{Sim00}) and optical and near-infrared observations 
   \citep{Bal00}. 
   Notwithstanding the observational and theoretical efforts, basic questions 
   connected to the mass accretion and ejection process are still to be 
   answered. First of all, it has still to be clarified what causes the end 
   of the accretion process, determining the final mass of a newly born 
   protostar. If, in the earliest evolutionary  phases,  winds and/or jets are 
   believed to play a fundamental role to expel excess angular momentum and 
   allow mass accretion, in a later phase these structures might evolve and be 
   responsible   for stopping the mass flow towards the protostar's surface. 
   Therefore the study of the protostellar jet dynamics and the mechanisms of 
   jet acceleration and collimation is of great importance.

   The earliest evolutionary phase of the forming YSO (Young Stellar Object), 
   the so-called Class 0 phase \citep{And94}, is characterized by the most
   active accretion-ejection phenomena. It is now believed that most of the 
   stellar mass is accreted during the Class~0 phase, while the mass accretion 
   rate drops rapidly during the subsequent Class~I phase. It is clear that these are 
   the most relevant phases for the investigation of the evolution of protostellar jets 
   and their possible relationship with the accretion process. To 
   characterize this relationship, it is essential to resolve jet structure 
   close to the central engine on a spatial scale of $\leq$ 10~AU. However, 
   the spatial resolution of $\sim$ 1000~AU provided by present millimeter 
   interferometer observations is insufficient to perform such a study. At 
   present, only Very Long Baseline Interferometry (VLBI) applied to water maser 
   lines allows us to explore the kinematics of the jet gas in the vicinity {\textemdash}
   down to sub-AU scales {\textemdash} of the accretion disk.

   Water masers at 22.2~GHz are well known to be associated with outflow
   phenomena and with the earliest evolutionary phases of star formation.
   Using the Nobeyama 45-m telescope, \citet{Fur01,Fur03} performed 
   a multi-epoch H$_2$O maser survey and revealed that maser activity is 
   high during the earliest stages of low-mass star formation, i.e. the
   Class~0 phase. The survey also showed that the maser activity fades 
   towards Class I, at the end of the main accretion and outflow phase, 
   and completely disappears after the pre-main sequence evolution 
   (Class~II), 
   when the star is optically visible and no significant accretion is 
   occurring.

   There are only a few published VLBI water maser studies towards
   low- and intermediate-mass young stellar objects \citep{Cla98, Fur00, 
   Pat00, Set02}.
   These VLBA studies indicated that the water masers trace predominantly 
   knot and shock structures at the base of and along the protostellar jets, 
   and can also emerge from parts of the protostellar disks.
   The main advantage of VLBA H$_2$O maser observations with respect to
   other techniques are that they: 1) probe the inner region of the disk/jet system 
   where the high extinction, especially in Class~0 objects, prevent 
   observations at shorter wavelengths; \ 2) measure proper motions and 
   line of sight velocities, and hence derive the 3-dimensional (3D) velocity 
   structure of the maser environment, with a time sampling 
   of only a few weeks, much shorter than the (typically) several 
   years required by 
   conventional high-resolution optical and near-infrared techniques (HST and 
   adaptive optics).

We have been using the VLBA to observe 22.2~GHz water maser
   emission towards a sample of low-mas YSOs in different evolutionary stages
   (from Class~0 to Class~I) with the final aim to study the evolution of the 
   disk/jet system as traced by the water masers. Our ultimate goal is to 
reconstruct the kinematic structure of the masing gas by means of 
milliarcsecond proper motion observations. 
Whenever possible we will use the phase-referencing technique, which allows
to measure absolute positions of the water maser features at each epoch.
In this paper we report the results of the first two sessions of 
    VLBA multi-epoch observations towards the cluster of low-mass YSOs in the Serpens 
    molecular core and towards RNO~15-FIR. For this latter source we were
able to obtain absolute proper motions.

\begin{table*}
\centering
\caption{VLBA Observations} \label{obs_table}
\begin{tabular}{|c|c|c|c|c|}
\hline
\multicolumn{1}{|c|}{Sources}& \multicolumn{1}{|c|}{First Run} & \multicolumn{1}{|c|}{Second Run} & \multicolumn{1}{|c|}{Third Run} & \multicolumn{1}{|c|}{Fourth Run} \\
 & \multicolumn{1}{|c|}{yr \ \ month \ \ day} & \multicolumn{1}{|c|}{yr \ \ month \ \ day} & \multicolumn{1}{|c|}{yr \ \  month \ \ day} & \multicolumn{1}{|c|}{yr \ \  month \ \ day} \\ 
\hline
Serpens SMM1 & 2003 April 1 & 2003 April 22 & 2003 May 19 & 2003 June 8 \\
Serpens SMM3 & 2003 November 21 & & & \\
Serpens SMM4 & 2003 November 21 & & & \\
RNO~15-FIR & 2003 October 19 & 2003 November 7 & & \\
\hline
\end{tabular}
\end{table*}

\section{Sample selection and strategy}

Water masers in low mass YSOs are on average much fainter than 
those observed in massive star forming regions. Additionally, the 
extensive 
single dish and VLA search and monitoring project towards low
mass young stellar objects  of \citet{Fur03} has clearly shown
that maser features are (as for high-mass stellar objects)
transient and highly variable. Many sources alternate quiescent and
active periods. These probably correspond to the formation, evolution and
fading of masing spots in the circumstellar regions.

\begin{figure*}
\centering
\includegraphics[width=15cm]{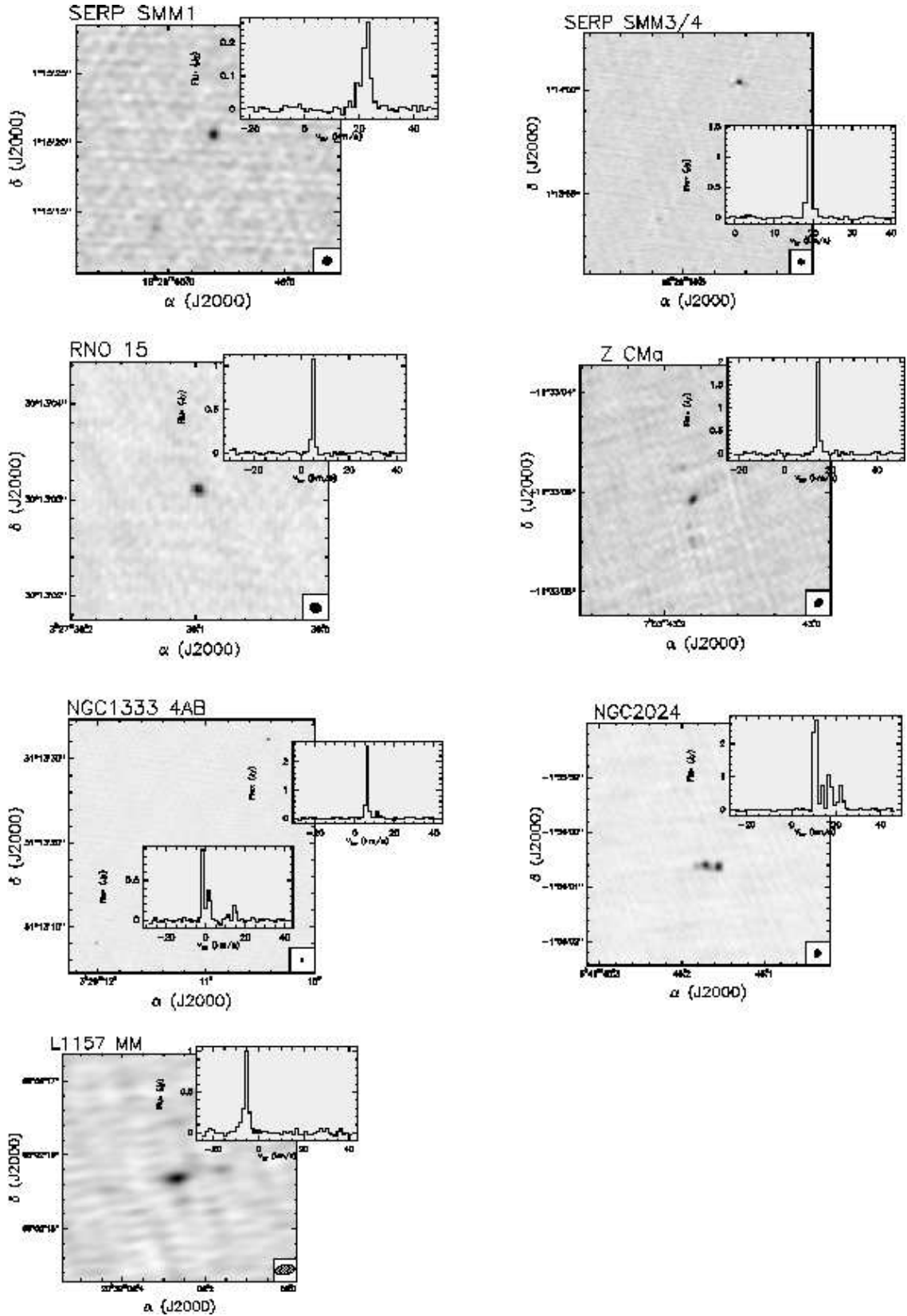}
\caption{Each plot shows the VLA map and spectrum of one source (whose name
is indicated
on top of the plot) detected with VLA snapshot observations prior to the VLBA sessions.}
\label{vla_fig}
\end{figure*}

In order to be able to efficiently use the VLBA time, we decided to 
perform snapshot surveys using the VLA of a large list of objects from the
survey of \citet{Fur03}. The selected objects span a wide range of bolometric
temperature \citep{Gre97}, corresponding to evolutionary stages from Class~0 to Class~I. All objects in our sample were detected at least in
one epoch by \citet{Fur03}. In March and September 2003 all
sources in our sample were observed and, consistent with our expectations, only 
a relatively small fraction of the sources were detected.
The sources detected
in our VLA snapshot survey are: L1157~MM, NGC1333~4AB, NGC2024, RNO~15-FIR,
Serpens SMM1~and~SMM3/4, Z~CMa. Except for NGC2024 (where more diffuse emission
appears), the VLA maps of these
sources show isolated maser spots, indicating that the water maser emission
associated with each YSO is distributed within an area of diameter of a few 
tenths of an arcsecond. Fig.~\ref{vla_fig} shows the VLA maps and spectra of all the detected sources.

The sources selected for the VLBA follow-up were the Class~0~and~I sources SMM1, 3, and 4 in
the Serpens molecular core and the Class~I source RNO~15-FIR. All the Serpens
sources are associated with molecular outflows and are among the brightest
members of the Serpens protocluster at millimeter wavelengths \citep{Tes98,Tes00}.
On the other hand,  RNO~15-FIR is a relatively isolated YSO, with a well detected 
bipolar outflow \citep{Dav97a}.



\section{Observations}

We used the VLBA (NRAO\footnote{The National Radio Astronomy Observatory (NRAO) 
is operated by Associated Universities, Inc., under cooperative agreement with 
the National Science Foundation.}) to observe the \(6_{16}-5_{23}\) H$_2$O maser 
line (at rest frequency 
22235.080~MHz) towards Serpens~SMM1 for four epochs each separated by about 
20 days in April-June 2003. A second session of VLBA multi-epoch observations
was not completed, and included a single epoch (November 2003) 
towards the two Serpens sources 
SMM3 and SMM4, and two epochs (separated by 19 days in October-November 2003)
towards RNO~15-FIR. Table~\ref{obs_table} lists the dates of the observing runs. 
In each observing run all ten antennas of the VLBA 
took part in the observations, except for the last epoch towards SMM1, when 
the Kitt Peak antenna was undergoing major maintenance and the Pie Town antenna
was replaced by a VLA antenna. 
Each epoch consisted of 7~hours of
integration on the target maser source(s), interrupted every 1-2~hours with 
a few-minute scan on strong continuum calibrators (3C345, J1751+0939,
J2005+7752 for the Serpens sources; 3C84, DA193, J0102+5824 for RNO~15-FIR)
for calibration purposes. At the time of our VLBA observations, no suitable
phase-reference calibrator was known within a distance of a few degrees from the 
Serpens sources. 
For the source RNO~15-FIR, we found an intense, compact 
continuum source sufficiently close (within 3\degr) to the maser target. 
In this case, we used a phase-referencing technique, which allows the derivation
of {\em absolute} positions and motions of the maser features.
The phase-reference source, J0336+3218, belongs to the list that
defines the International Celestial Reference Frame (ICRF), and its
absolute position is known with high accuracy ($\leq 2$~mas).  
The 4~cm VLBA map of J0336+3218 shows an elongated structure 
($\approx$10~mas in size) with a peak flux density of 0.5~Jy~beam$^{-1}$.

Both circular polarizations were recorded using 16~MHz bandwidth for 
the sources SMM3, SMM4 and RNO~15-FIR, and 4~MHz bandwidth for the source SMM1. 
The data were processed at the VLBA correlator in Socorro (New Mexico, USA) 
using either 512 (for the 4~MHz bandwidth) or 1024 spectral channels (for
the 16~MHz bandwidth), corresponding to a velocity resolution of 0.1~km~s$^{-1}$ 
and 0.2~km~s$^{-1}$, respectively.

Data reduction was performed using the NRAO AIPS package, following the 
standard procedure for VLBI line data. Total power spectra of the strong 
continuum calibrators were used to derive the bandpass response of each 
antenna. Amplitude calibration was performed using the information on the 
system temperature and the gain curve provided automatically by each antenna 
to the VLBA correlator. 

For each observing epoch, a single scan of a strong calibrator was used to 
derive the (time-independent) single-band delay  and the 
phase offset between the two polarizations.After removing these major 
instrumental effects, all calibrator scans were fringe-fitted to determine the 
residual (time-dependent) delay and the fringe rate. In all cases the 
residual delays (always $\leq$ 15~ns) and fringe rates (always $\leq$ 5~mHz) were 
found to be sufficiently small not to require the maser line data to be 
further corrected for these minor instrumental effects.  

For each source and epoch, the visibilities of all the 
spectral channels were referred in phase to the velocity channel of maximum 
emission in the total power spectra.
The selected (phase-)reference features were detected with high signal-to-noise 
ratio (SNR) at each observing epoch, and, after imaging that velocity channel, 
we verified that they always exhibited a simple spatial structure consisting 
of a single, almost unresolved maser spot.  The visibilities of the reference 
channel were fringe-fitted to find the residual fringe rate produced both by differences 
in atmospheric fluctuations between the calibrators and the maser source, and by 
errors in the model used at the correlator. After correcting for the residual 
fringe rate, the visibilities of the reference channel were self-calibrated 
to correct for rapid time variations of atmospheric path lengths and to remove 
any possible effect induced by extended spatial structure. The self-calibration
process was effective in reducing the RMS noise level, $\sigma$, on the reference channel 
map by a factor of \ 2 -- 4 (depending on the observing epoch). 
Finally, the corrections 
derived from the reference channel were applied to the visibility data of all spectral 
channels. For the source RNO~15-FIR, the absolute position of the reference 
maser channel was derived applying the standard phase-referencing technique. The
data of the phase-reference calibrator were fringe-fitted and the derived phase
solution applied to the reference maser channel before imaging.

At first, by producing tapered, channel{\textendash}averaged maps, we searched 
for maser emission over the whole velocity range where signal was visible in the 
total-power spectra, and over a sky area of 
$(\Delta \alpha \ cos\delta \times \Delta \delta) \, 2\farcs5 \times 2\farcs5 $ 
centered on the reference feature. For each source (and epoch), we found maser 
emission centers to be distributed within an area with a maximum distance of 
$0.4\arcsec$ from the reference feature. The region of detection of maser 
emission was then imaged at full angular and velocity resolution, producing sensitive, 
naturally weighted maps.  The CLEAN beam was an elliptical Gaussian with a 
FWHM size, slightly varying from source to source, of 1.2 -- 1.5~mas along 
the major axis and \ 0.4 -- 0.6~mas along the minor axis. In each observing 
epoch, the RMS noise level on the channel maps, $\sigma$, is close to the 
theoretical thermal value, 3~mJy~beam$^{-1}$, for channels where no signal is 
detected, and increases to 10~mJy~beam$^{-1}$ for channels with the strongest 
components.

Every channel map was searched for emission above a conservative 
detection threshold (in the range 5{\textendash}10~$\sigma$), and the detected maser spots 
have been fitted with two-dimensional elliptical Gaussian, determining  
position, flux density, and FWHM size of the emission. Hereafter, we use the
term of ``spot'' to refer to compact maser emission in a single channel map 
and the 
term of ``feature'' to indicate a collection of spectrally and spatially 
contiguous maser spots. Contour plots of every channel map with relevant 
maser emission were produced in order to check whether the number of fitted
Gaussian components corresponded with the number of ``spots'' visible in the
image. A maser feature was considered real if it was detected 
in at least three contiguous channels, with a position shift of the intensity 
peak from channel to channel smaller than the FWHM size.

The uncertainty in the relative positions of the maser spots was estimated 
using the expression
\begin{equation}   
\Delta \theta = \frac{\sigma }{2 \, I } \; FWHM \,,
\end{equation}
where $FWHM$ is the full width at half maximum size of the spot
without deconvolution, $I$ is the peak intensity and
$\sigma$ is the no-signal RMS of the image \citep{Rei88}.
By using this formula, for most of the spots the positional uncertainty was 
found to be of the order of $\approx$1 -- 10~$\mu$as.

\section{Results}

\subsection{H$_2$O maser spectra}

\begin{figure*}
\centering
\includegraphics[width=10cm]{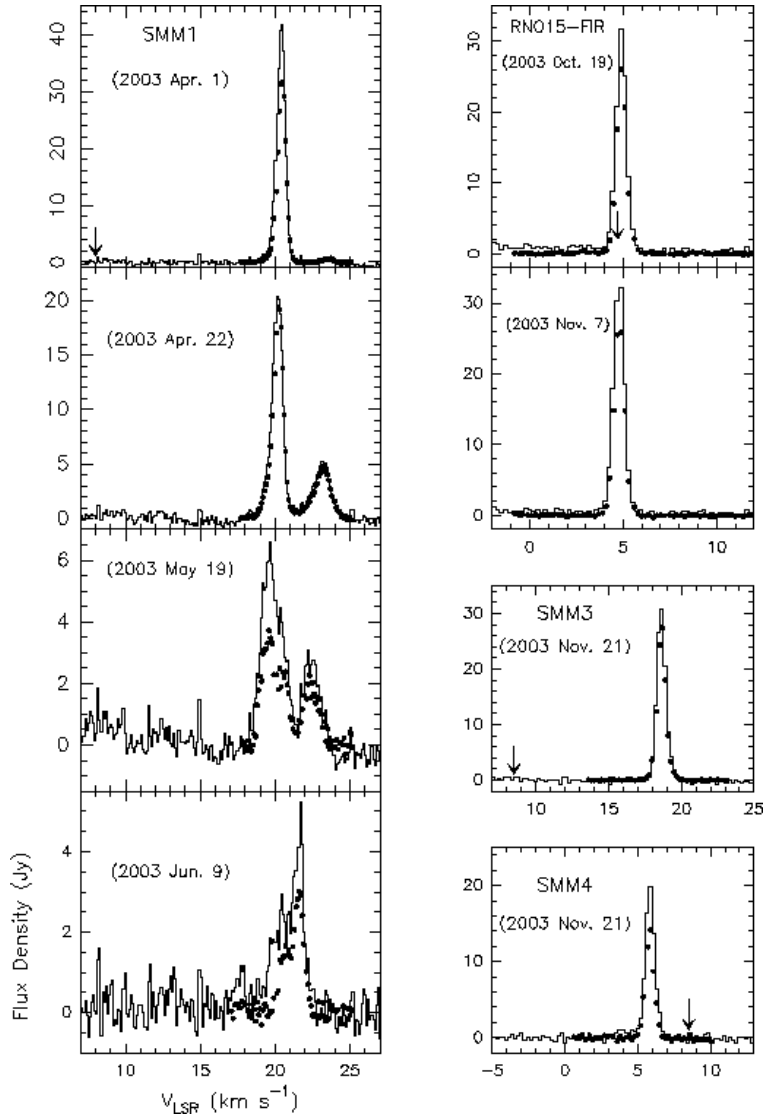}
\caption{Comparison of the Pie Town total power spectrum (histogram plot) 
with
the integrated flux densities of the VLBI channel maps (dotted plot). 
The panels
on the left column show the data from the four observing epochs of
the source Serpens SMM1. The four panels on the right column, going from top 
to the bottom, show data from the two observing epochs of
the source RNO~15-FIR, and from the single epochs towards the Serpens sources 
SMM3 and SMM4. For each source, a black arrow indicates the systemic velocity, as derived by thermal 
molecular observations \citep{Hog99, Dav97a}.}
\label{spec}
\end{figure*}

Fig.~\ref{spec} shows, for each source and observing epoch, the 
comparison between the Pie Town total-power spectrum and the 
integrated flux densities of the VLBI channel maps.  
We note that the difference between the total-power flux and 
the flux recovered in the imaged VLBI field of view is always within the 
amplitude calibration errors of $\pm$~20 -- 30\%. This suggests that our observations are
consistent with the assumption that all the maser emission within the 
VLBA primary beam is concentrated in the regions around the sources selected
for correlation of the data.

Figure~\ref{spec} provides evidence that the maser emission peak towards the 
source SMM1 has decreased from $\approx$40~Jy to $\approx$4~Jy over a time
span of two months. On the contrary, the maser spectrum towards the source
RNO~15-FIR does not show any significant variation over $\approx$20~days.  

We have compared the maser spectra in our VLA and VLBA observations with the 
results presented by \citet{Fur03}. All the sources selected for the
VLBA follow-up 
were highly variable in the single dish monitoring program,  and the  
spectral complexity and the intensity of the features was very different
from source to source. 

{\it RNO15}: The single dish observations showed only a single faint feature
at only one epoch. Our observations reveal a single spectral feature at 
approximately the same velocity as the single dish data from January 1999.

{\it Serp SMM1}: The single dish spectra show a complex and highly variable
structure \citep{Fur03}. In our VLA and VLBA observations we only detect a
group of features near V$_{LSR}\approx 20$~km~s$^{-1}$, while the strong component
near $\approx$10~km~s$^{-1}$ detected in the single dish spectra of June 1999 was not
detected in 2003.

{\it Serp SMM3/4}: These two sources could not be resolved in the \citet{Fur03}
single dish monitoring. The Nobeyama data show only a faint feature detected
in January 1999 near V$_{LSR}\approx 0$~km~s$^{-1}$.  Our VLA observations 
show two distinct groups of maser features, emitting at V$_{LSR}$ 
$\approx$20~km~s$^{-1}$ towards SMM3, and at V$_{LSR}$$\approx$6~km~s$^{-1}$ 
 towards SMM4.

\begin{table*}
\centering
\caption{Maser feature parameters} \label{features}
\begin{tabular}{ccccccccccc}
& & & & & & &  & & & \\
\hline\hline
\multicolumn{1}{c}{Source} & \multicolumn{1}{c}{Feature} & \multicolumn{1}{c}{$V_{\rm LSR}$} &
\multicolumn{1}{c}{$F_{\rm int}$} &  &
\multicolumn{1}{c}{$\Delta \alpha$} & \multicolumn{1}{c}{$\Delta \delta$} &
& \multicolumn{1}{c}{$V_{\rm x}$} & \multicolumn{1}{c}{$V_{\rm y}$} &
\multicolumn{1}{c}{$V_{\rm mod}$} \\
\multicolumn{1}{c}{name} & & \multicolumn{1}{c}{(km s$^{-1}$)} &
\multicolumn{1}{c}{(Jy)} &   &
   \multicolumn{1}{c}{(mas)} & \multicolumn{1}{c}{(mas)} &
   & \multicolumn{1}{c}{(km s$^{-1}$)} & \multicolumn{1}{c}{(km s$^{-1}$)} &
   \multicolumn{1}{c}{(km s$^{-1}$)} \\
\hline
& & & & & & &  & & & \\
 SMM1 &  1 &  20.0 & 13.96 & &     53.80 (0.34) &     53.22 (0.20) & &    17.8 (4.1) & 26.5 (3.5) &    31.9 (3.7) \\
 &  2 &  20.3 &  4.68 & &     53.19 (0.04) &     53.50 (0.08) & &       &   &       \\
 &  3 &  20.2 &  0.14 & &     52.98 (0.05) &     54.38 (0.11) & &       &   &      \\
 &  4 &  19.7 &  0.50 & &     51.59 (0.04) &     53.62 (0.07) & &    {\it 36.1 (2.7)$^{\dag}$} & {\it 22.1 (3.0)} &  {\it  42.3 (2.8)} \\
 &  5 &  19.4 &  0.16 & &     65.75 (0.04) &     50.88 (0.08) & &    {\it 25.0  (0.4)} & {\it 12.7 (1.1)} &   {\it 28.1 (0.6)} \\
 &  6 &  20.9 &  0.32 & &     62.87 (0.04) &     51.63 (0.07) & &    26.7 (1.1) & 25.1 (1.2) &    36.7 (1.1) \\
 &  7 &  23.6 &  0.91 & &     --0.22 (0.04) &      0.95 (0.07) & &  {\it --17.4 (3.6)} & {\it 1.7 (3.4)} & {\it 17.4 (3.6)} \\
 &  8 &  22.5 &  1.47 & &      0 &      0 & &     0 & 0 &     0 \\
 &  9 &  20.8 &  0.10 & &     57.39 (0.09) &     53.49 (0.09) & &      &  &      \\
 & 10 &  22.7 &  2.00 & &     --2.20 (0.13) &      2.12 (0.15) & &      &  &      \\
 & 11 &  18.0 &  0.16 & &     47.98 (0.09) &     53.40 (0.09) & &      &  &      \\
 & 12 &  20.3 &  1.57 & &     63.71 (0.21) &     53.82 (0.25) & &      &  &      \\
 & 13 &  23.0 &  0.36 & &      2.70 (0.22) &     --3.22 (0.27) & &      &  &      \\
& & & & & & &  & & & \\
 SMM3 &  1 &  18.6 & 23.16 & &      0 &      0 & &      &  &     \\
 &  2 &  19.5 &  0.70 & &     --0.55 (0.05) &      0.30 (0.13) & &      &  &      \\
 &  3 &  18.8 &  1.48 & &     --0.76 (0.05) &      0.00 (0.14) & &      &  &      \\
 &  4 &  18.1 &  0.37 & &     --1.22 (0.04) &     --0.67 (0.13) & &      &  &      \\
 &  5 &  18.8 &  1.09 & &     --1.79 (0.10) &      0.26 (0.20) & &      &  &      \\
 &  6 &  19.5 &  0.09 & &     --2.37 (0.04) &      0.46 (0.13) & &      &  &    \\
& & & & & & &  & & & \\
 SMM4 &  1 &   5.9 &  9.56 & &      0 &      0 & &      &  &      \\
& & & & & & &  & & & \\
 RNO~15-FIR &  1 &   4.9 & 25.06 & &      0 &      0 & & {\it  --22.8 (13.6)} & {\it --16.2 (13.6)} &  {\it  27.9 (13.6)} \\
 &  2 &   5.0 &  0.90 & &     --0.51 (0.31) &     --0.32 (0.32) & &   {\it --32.5 (14.3)} & {\it --17.3 (14.2)} &  {\it  36.8 (14.3)} \\
 &  3 &   2.9 &  0.21 & &     --2.47 (0.31) &     --1.89 (0.32) & &    {\it --4.4 (14.4)} & {\it --4.5 (14.6)} &   {\it  6.3 (14.5)} \\
 &  4 &   5.1 &  0.31 & &      0.78 (0.32) &      0.42 (0.34) & &      &  &     \\
& & & & & & &  & & & \\
\hline
\end{tabular}
\begin{flushleft}
${\dag}$ The italic character is used to indicate tentative values of proper motion components for  features observed either at only two epochs or with a very large uncertainty in the  direction of motion.
\end{flushleft}
\end{table*}

\subsection{Maser feature identification and proper motion}

For each observed source, Table~\ref{features} lists the
maser features' parameters, derived by fitting an elliptical Gaussian to the
intensity distribution of the maser spots contributing to the features' emission
at different velocity channels. Col.~1 indicates the source name, and the corresponding
maser features are denoted with the label numbers given in Col.~2. 
Cols.~3~and~4 report respectively the line of sight velocity, $V_{\rm LSR}$, 
and the integrated flux density, $F_{\rm int}$, of the highest-intensity
channel, with values averaged over the observational epochs for the time-persistent
features. Cols.~5~and~6 report the feature positional offset (of the
first epoch of detection) calculated with respect to the reference feature.
The positional offset of a given feature is estimated from the 
(error-weighted) mean position of the contributing maser spots. The
 bracketed numbers are the positional uncertainties, evaluated by taking the 
weighted standard deviation of the spot positions.

\begin{figure*}
\centering
\includegraphics[width=11.5cm]{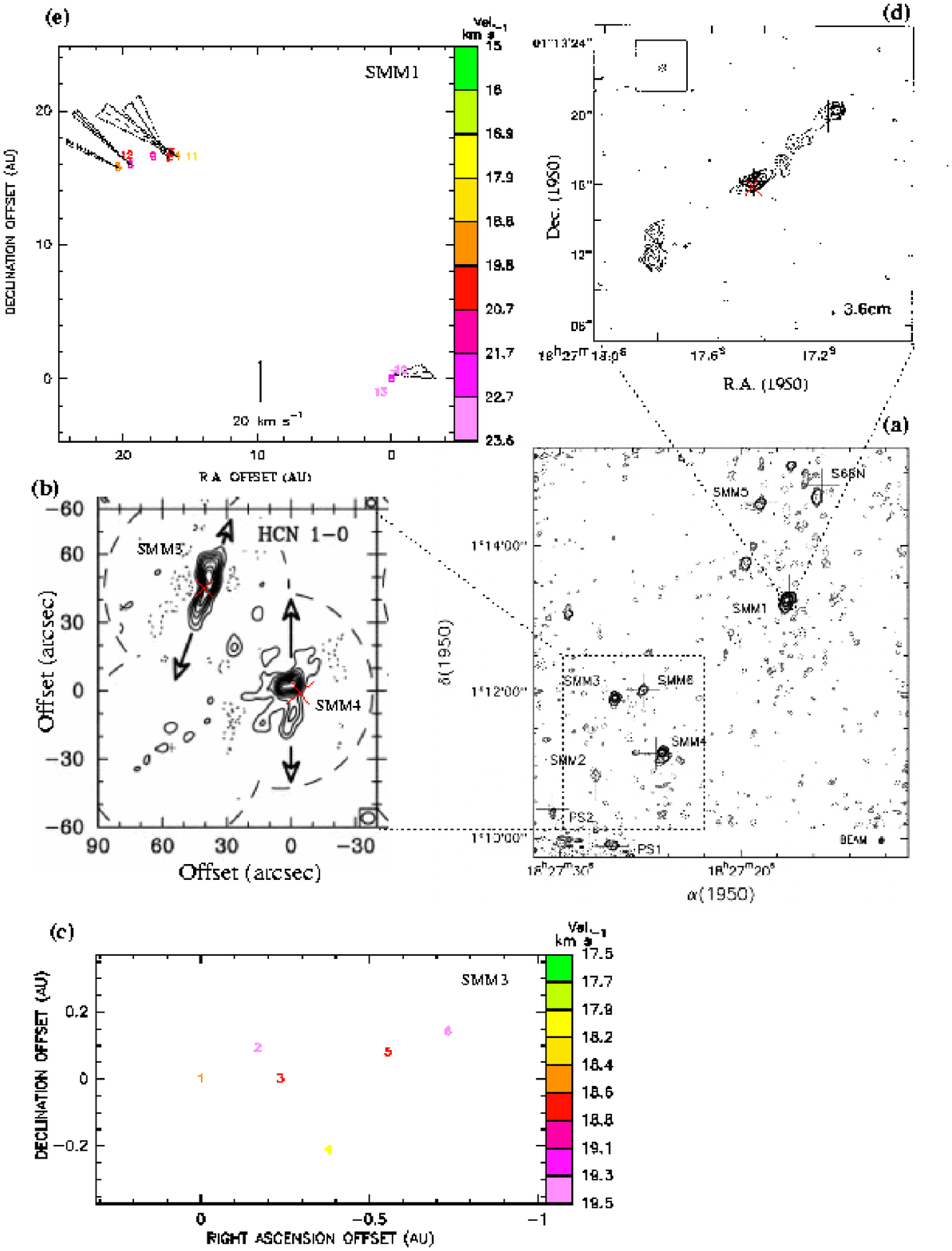}
\caption{VLBA results towards the Serpens protocluster. \ (a) OVRO 3~mm continuum mosaic from \citet{Tes98}; 
\ (b) OVRO HCN \ J $=$1 $\to$ 0
image towards SMM3 and SMM4 \citep{Hog99}. The arrows indicate the position angles of larger
scale $H_{2}$ jets. The red crosses indicate the positions of the water masers
detected with the VLA. \ (c) Spatial distribution of water masers 
towards SMM3 from our single-epoch VLBA observation. Each
maser feature is identified with the label number given in Col.~2 of 
Table~\ref{features}. Different colours are used to distinguish the line of 
sight velocities of the features, according to the colour-velocity conversion
code shown on the right-hand side of the panel. The positional coordinates are
relative to the reference feature (with label number ``1'') and are given in AU. \ (d) VLA 3.6~cm map of the radio jet towards SMM1 obtained
by \citet{Cur93}. The red cross indicates the VLA water maser position. \ (e) Spatial
distribution and proper motions of the water maser features towards SMM1 from
our four-epoch VLBA observations. Feature positions and line of 
sight velocities are indicated with the same notation as in panel ''c''.
The triangles around the proper motion vectors represent the uncertainty in 
amplitude and orientation.
The positions and proper motions are relative to the feature with label number ''8''. The amplitude scale of the proper motions is given at the
bottom of the panel.}
\label{serpens}
\end{figure*}

As explained in Section~3,
in order to increase the signal-to-noise ratio, for each source 
and epoch the fringe-fit and self-calibration solutions
were determined using the strongest spot (always belonging to 
feature ''1'',except for the fourth epoch towards Serpens SMM1, where 
the strongest emission comes from feature ''8''), and images 
were made referencing the visibilities of all the channels to this spot.
In Table~\ref{features}, positions of each feature of the Serpens sources 
SMM3 and SMM4 and of RNO~15-FIR, are listed relative to feature ''1'' for
each source.   In the case of the Serpens source SMM1, maser 
features are concentrated in two separated clusters
(directed along a northeast-southwest direction; see discussion in 
Section~4.3). Since most of the proper motions are measured in the 
NE cluster, in order to show the relative motion of the two (NE and SW) 
clusters of maser spots, Table~\ref{features} (and Figure~\ref{serpens}) 
present results with positions (and proper motions) referred to a 
persistent feature (label number ''8'') of the SE cluster.

Taking an average of the FWHM sizes along the major and minor axes, the VLBA 
synthesized beam varied from source to source and from epoch to epoch from
\ 0.6~mas to 1.2~mas.  In each source, and for each detected feature,
the deconvolved size of the strongest (and best characterized) maser spots 
is always found to be smaller than the observing beam size. A few
($\leq$ 10\%) of the weaker spots have deconvolved sizes slightly larger than
the beam, but these measurements are uncertain owing to 
the low signal-to-noise ratio. Since our sources have estimated
distances between 300~and~400~pc, upper limits to the spot sizes in the range
 0.2 -- 0.4~AU can be derived. 

Criteria for cross-epochal matching of features were based on several
 considerations and requirements. 
 We matched features appearing at two non-consecutive epochs, as well.
Assuming maser emission originates in shocked layers of gas, since the conditions for maser
action are highly variable, during the lifetime of a given shock distinct
episodes of maser emission can occur. Then, in particular for the weakest
maser features, it is plausible that they can fade away at a given epoch
and reappear subsequently.
The features' (relative or absolute) proper
motions are searched for velocities less than 250~km~s$^{-1}$, which,
for a time baseline of $\approx$2~months and a source distance of 300~pc, 
requires that positions of features corresponding over epochs 
have to be closer than 5~mas. If, as in the case of the source Serpens SMM1, 
maser features 
are found to be grouped within clusters of given shape and size, 
establishing cross-epochal correspondence may be facilitated by the 
persistence 
of the cluster geometry. Finally, for features observed at three (or more) 
epochs, a good
check of the consistency of the cross-epochal matching is that the features
 move along smooth (linear or curved) trajectories.

\begin{figure*}
\centering
\includegraphics[width=17cm]{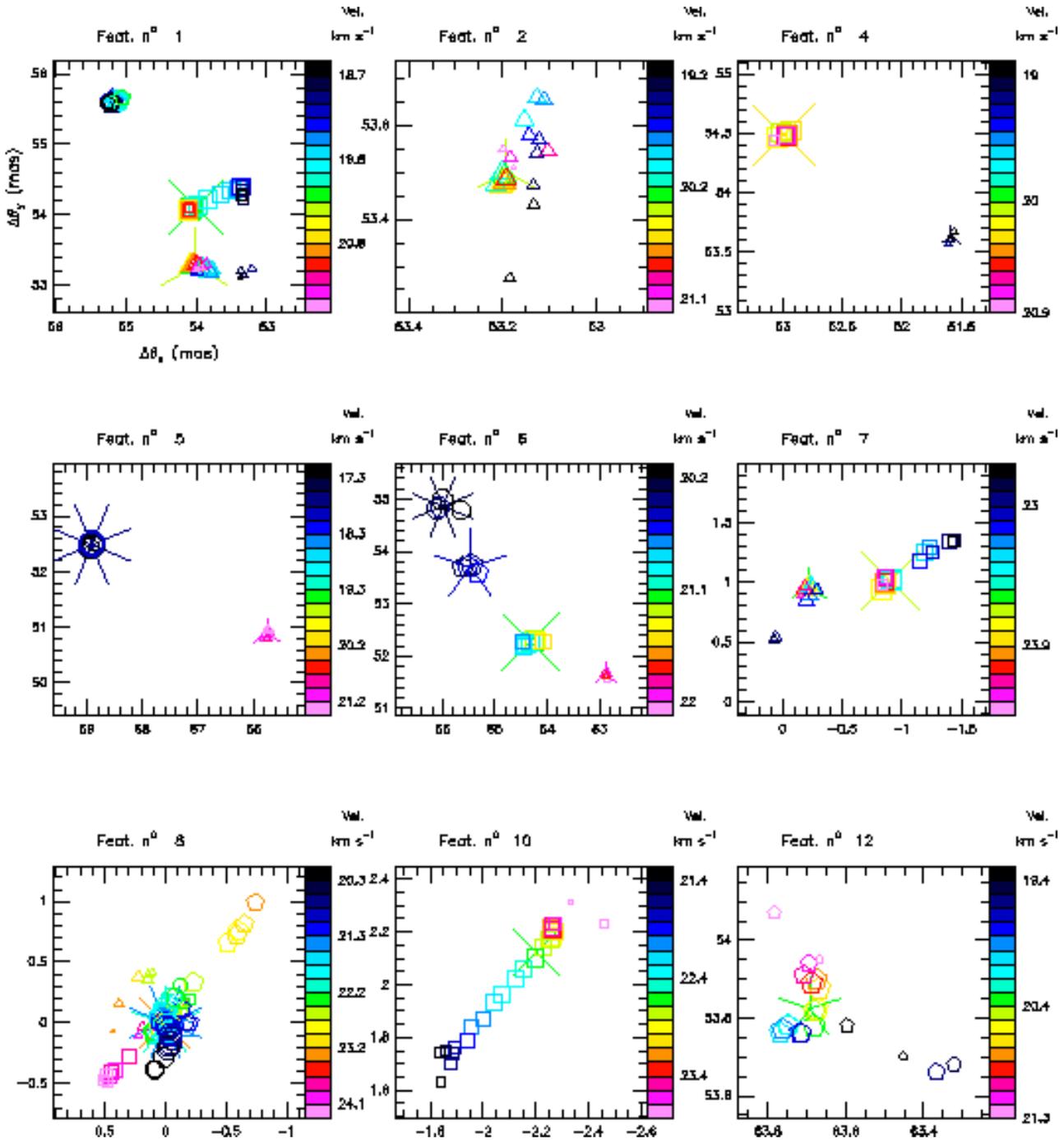}
\caption{Spatial and velocity distribution of features' spots in Serpens SMM1. 
Each plot 
 refers to a different maser feature (whose label number is indicated on top 
of the plot) and reports the positions and the line of sight velocities of the
spots contributing to the feature emission at various observing epochs.
To distinguish the epochs, symbols with ever larger numbers of sides are
used (triangle
for the first epoch, squares for the second epoch, etc.). Vertex-connected 
polygons give the feature positions (i.e., the 
error-weighted mean positions of the contributing spots) at different epochs. The symbol colours,
from heavy blue to magenta, indicate increasing values of the spot or 
feature V$_{LSR}$ across the emission range of all the epochs. The colour-velocity 
conversion code is shown on the right-hand side of the plot.
Each feature V$_{LSR}$ is determined from the intensity-weighted mean 
V$_{LSR}$ of the contributing spots. Symbol sizes scale logarithmically 
with the spot or feature intensity (the feature intensity, at each epoch, 
is taken equal to the strongest spot intensity).}
\label{sp_epo_smm1}
\end{figure*}

\begin{figure*}
\centering
\includegraphics[width=18cm]{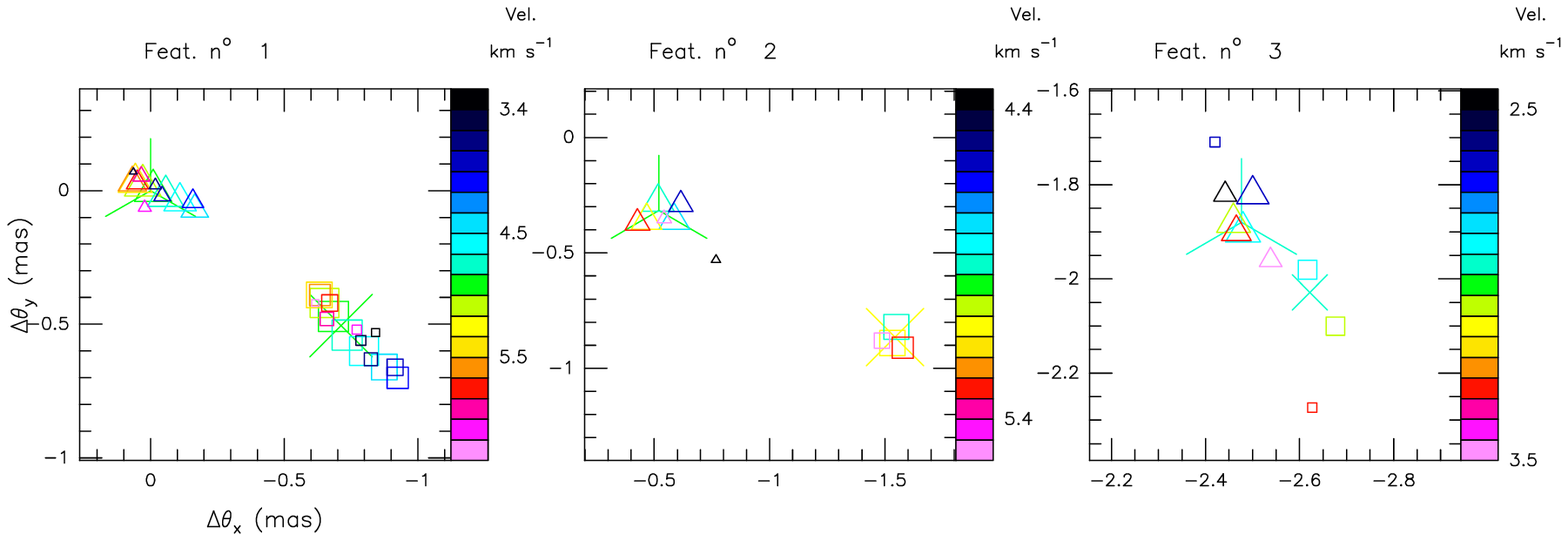}
\caption{Spatial and velocity distribution of features' spots in RNO~15-FIR. 
 Symbol shapes, colours and sizes have the same meaning as in 
 Figure~\ref{sp_epo_smm1}.}
\label{sp_epo_rno15}
\end{figure*}

For most of the features observed towards the sources Serpens SMM1 and 
RNO~15-FIR, respectively,
Figures~\ref{sp_epo_smm1}~and~\ref{sp_epo_rno15} show the spatial and
velocity distribution of the spots contributing to the features' emission.
In both sources, for the persistent features, spots at different epochs are 
clearly separated 
in the sky-plane, with a monotonic shift in both spatial coordinates when the 
emission is observed for more 
than two epochs (for features labeled ''1'' and ''6'' of SMM1). 
 In SMM1, feature ''5'', among the weakest features (see Table~\ref{features}), 
is detected only at the first and last epoch.

\begin{figure*}
\centering
\includegraphics[width=18cm]{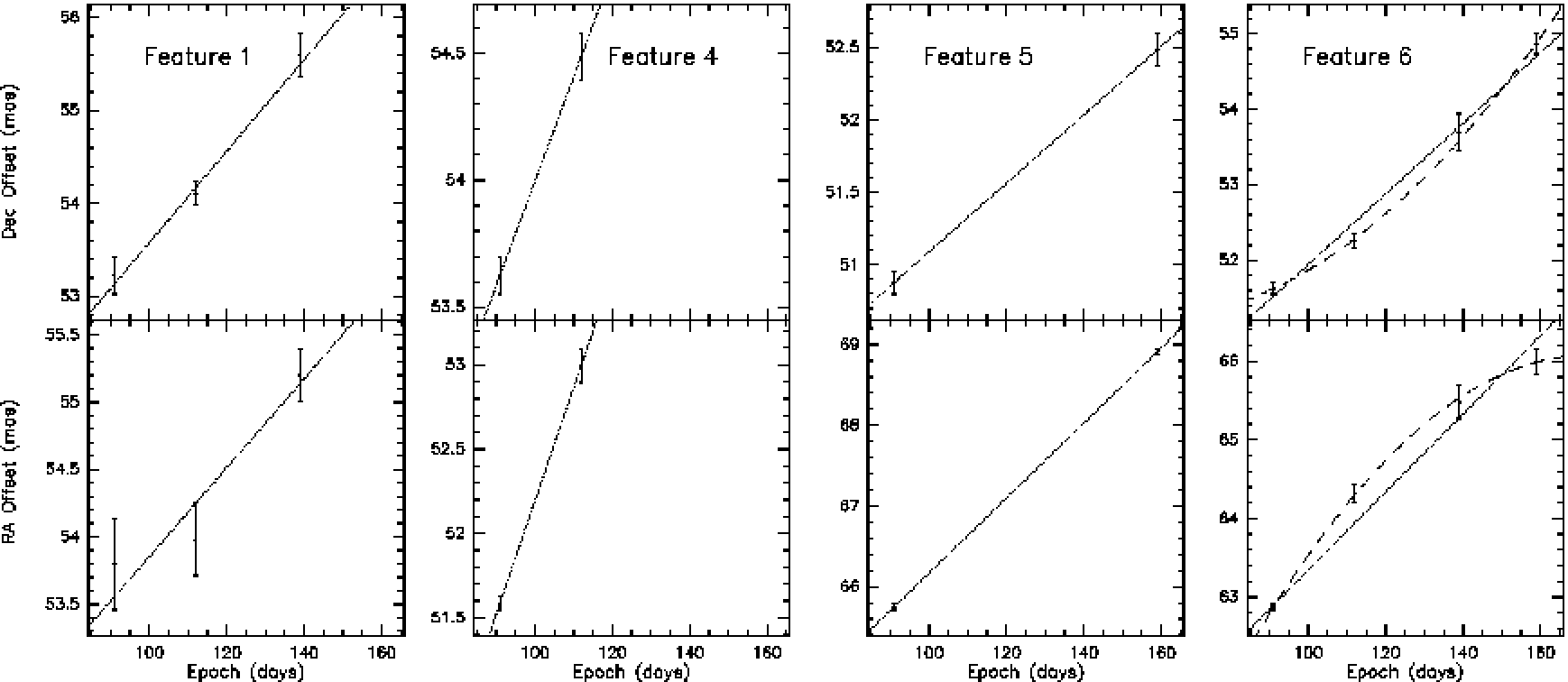}
\caption{Relative proper motion of four persistent features observed towards the Serpens 
source SMM1. Each plot refers to a different feature (indicated with the same
labeling as Table~\ref{features}) and shows the variation of the R.A. 
(lower panel) and Dec (upper panel) offsets across the observing epochs.
Features' positional offsets are relative to the feature ''8'' (see 
Table~\ref{features}). In each panel, the dotted line shows 
the linear least-squares fit of positions vs time. For feature ''6'', a 
second-order polynomial fit has been also performed, and the result is
shown by the dashed line.}
\label{prop-mot}
\end{figure*}

Figure~\ref{prop-mot} presents plots of the variation of the 
position with
time for four persistent features observed toward the source Serpens SMM1. 
The proper motions have been calculated by performing a (error-weighted) linear
least-squares fit of the positional offsets with time. The derived proper
motions are a measure of the average motion of the features over the epochs.
Tentative values of the proper motions are also calculated for features 
observed at only two epochs.
For features observed at three or four epochs (''1'' and ''6''), Figure~\ref{prop-mot} 
indicates that, given the positional 
uncertainties, a constant velocity motion is a reasonable approximation of
their real motion. Particularly interesting, however, is the case of 
feature ''6'', for which compelling evidence of acceleration does exist.
Looking at Figure~\ref{sp_epo_smm1}, one can note that this feature 
appears to move along a curved rather than a linear trajectory, and 
from Figure~\ref{prop-mot} one sees that a second-order polynomial fits
the variation of positions with time better than a linear fit. In Section~5.3,
the motion of feature ''6'' is discussed within the scenario
proposed to explain the water maser kinematics in Serpens SMM1.

Cols.~7,~8~and~9 of Table~\ref{features} report 
the projected components along the R.A. and DEC axis, and the 
absolute value of the derived proper motions, respectively.
The numbers in italics refer to features observed at only two epochs.
The bracketed numbers are the formal errors of the linear least-squares fit.
From Table~\ref{features}, and Figures~\ref{sp_epo_smm1},
\ref{sp_epo_rno15}~and~\ref{prop-mot}, it is clear that 
nearby features are always found to move with similar velocities, both in
magnitude and direction. This result gives us some confidence that the
the derived proper motions are correct, also for features 
observed at only two epochs.

In the following subsections, the results obtained towards the sources 
in the Serpens core and towards RNO~15-FIR are presented separately.

\subsection{Serpens core}

For the Serpens molecular core, Fig.~\ref{serpens} shows a collage of previous 
interferometric observations with our VLBA maps. 
Fig.~\ref{serpens}a is the 99~GHz map obtained by \citet{Tes98} using the OVRO
(Owens Valley Radio Observatory) interferometer. Over a region of size about \ 
 $ 5^{'} \times 5^{'}$ \ many 3~mm condensations are visible, suggesting that
the process of fragmentation and collapse of the molecular cloud is in a very
active phase. Interferometric molecular line observations \citep{Hog99} 
reveal that typical condensation masses are in the range 1 -- 10~M$_{\odot}$ and
gas temperatures are $\leq$ 50~K. Most of the massive 
condensations visible in the \citet{Tes98} 3~mm map are thought to 
harbour a low-mass protostar accreting matter from the surrounding molecular 
envelope. 

Fig.~\ref{serpens}b shows the OVRO HCN \ J $=$1 $\to$ 0 map by 
\citet{Hog99} towards the two Serpens YSOs SMM3 and SMM4. In both
sources, the HCN emission
is extended along the direction (indicated with an arrow) of a larger scale
H$_{2}$ jet, and likely traces ambient gas shocked by the fast 
outflow emerging from the protostar. The red crosses give the position 
(accurate within $\approx$0$\farcs$1) of the 22.2~
GHz water masers observed in this region using the VLA just a few
days before the VLBA observations. The positional accuracy of the HCN map,
$\sim$ 1$^{''}$, is sufficient to establish that, in both sources, the
water maser emission emerges close to the HCN peak.

A single water maser feature has been 
detected in SMM4, blue-shifted by a few \ km~s$^{-1}$ with respect to the 
LSR velocity of the ambient gas, 
$\approx$8.5~km~s$^{-1}$.
Fig.~\ref{serpens}c reports the 
spatial and line of sight velocity distribution of the water maser features
detected towards SMM3. 
The maser features display an elongated, albeit very small ($\leq$ 1~AU in 
size),
spatial distribution, lying approximately parallel to the  
R.A. axis. It is worth noting that, although the
angular scale is much larger, the molecular outflow observed towards SMM3 
is oriented close to the Dec axis (see. Fig.~\ref{serpens}b), and therefore
at a large
angle from the axis of the water maser distribution. The maser
features detected in SMM3 are strongly red-shifted, with a line of sight 
velocity 10 -- 12~km~s$^{-1}$  higher than the LSR velocity of the
ambient gas.
      
Fig.~\ref{serpens}d shows the radio jet observed towards SMM1 by \citet{Cur93}
using the VLA at 3.6~cm. It consists of three main components aligned
along a southeast-northwest direction (P.A. = $-$53\degr), with the central and
northwest components connected by a bridge of weaker emission. The central component appears
elongated, having a deconvolved size of \ 200 $\times$ 
$\leq$~60~AU, with a position angle close to that of the radio jet.
The radio continuum emission of the northwest jet component is thought to originate
from shock waves produced by bullets emitted by the YSO into the surrounding,
dense ambient material. The red cross indicates the VLA position of the 
22.2~GHz
water masers, located close to the emission peak of the central component.
It should be noted that the separation between the maser and the
continuum peak (0$\farcs$3--0$\farcs$4) is of the same order as the positional
accuracy of the VLA maser (0$\farcs$1) and continuum (0$\farcs$2) maps, and
therefore, with present data, we cannot say whether or not it is significant.

Fig.~\ref{serpens}e shows the water maser map derived with four epochs of VLBA 
observations towards SMM1. Thirteen water maser features are detected and for
five of them, observed at two or more epochs, (relative) proper motions are
measured. Maser features are concentrated in two small clusters (size $\leq$ 5~AU), 
separated on the sky by $\approx$25~AU along a 
northeast-southwest direction (P.A. $\approx$ 50\degr ). The two clusters 
appear 
to  move away from each other, as indicated by the (relative) transverse 
velocities measured for features in the northeastern cluster, having similar
values (30 -- 40~km~s$^{-1}$) and
consistent motion away from the southwestern cluster.
All the maser features are strongly red-shifted with respect to the LSR velocity of the
ambient gas, by 10 -- 12~km~s$^{-1}$ for the masers in the 
northeastern cluster, and by  14 -- 15~km~s$^{-1}$  for those in the 
southwestern cluster.

\begin{figure*}
\centering
\includegraphics[width=9cm]{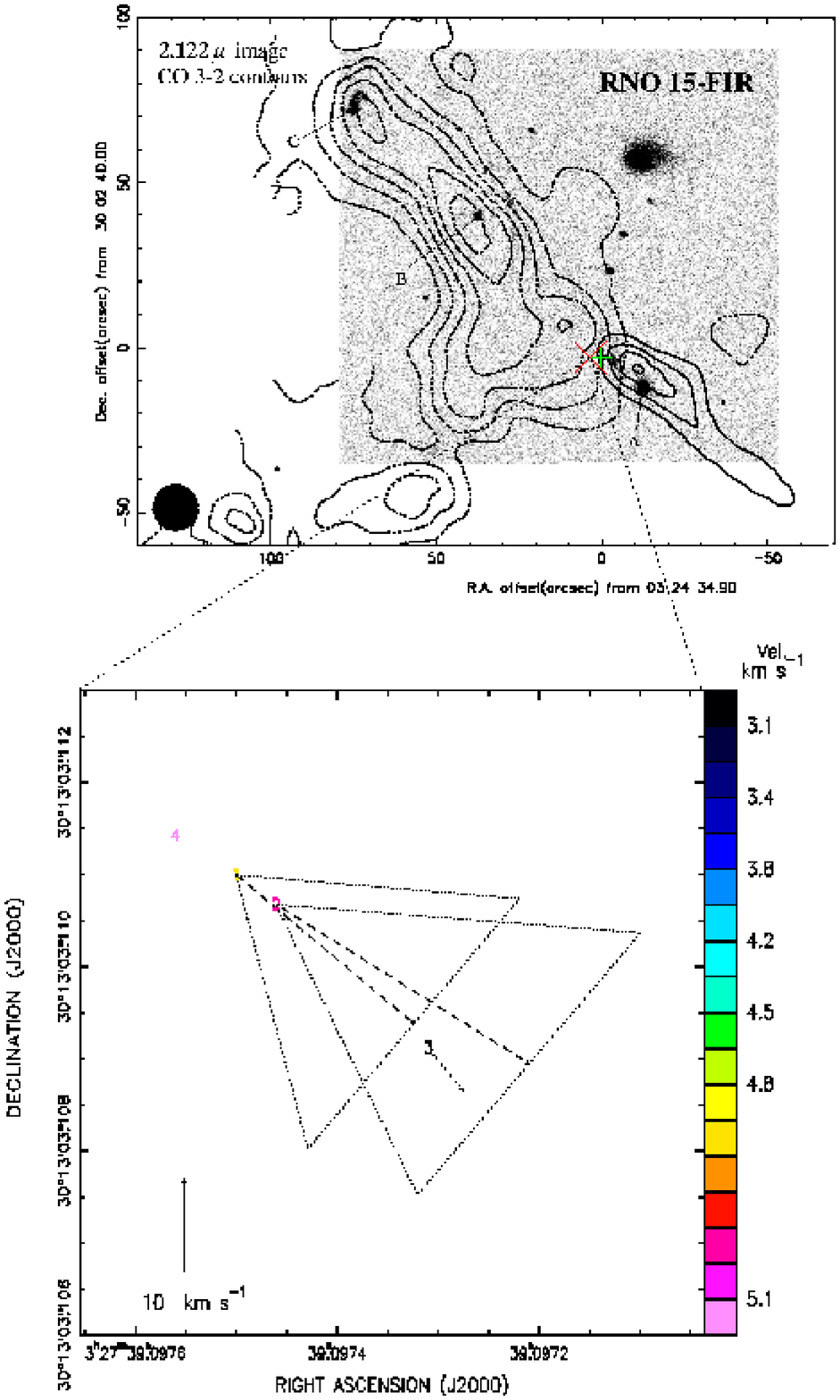}
\caption{VLBA results towards RNO~15-FIR. The top panel shows a H$_{2}$
2.12~$\mu$m gray scale image of the RNO~15-FIR outflow region \citep{Dav97a}
with a contour map of the CO $3~\to~2$ integrated intensity measured by 
\citet{Dav97b} using the James Clerk Maxwell Telescope (JCMT) overlaid.
The blue-shifted flow lobe is shown with continuous lines, and the red-shifted lobe
with dashed lines. The three brightest compact H$_{2}$ knots are denoted 
with the letters ''A'', ''B'' and ''C''. The green cross marks the IRAS
position of RNO~15-FIR. The red cross gives the absolute position of the
22.2~GHz water masers as determined from our phase-referenced VLBA observations.
\  The bottom panel shows the water maser map derived with two epochs of
 phase-referenced VLBA observations. Feature positions and line of
sight velocities are indicated with the same notation used for the water
maser maps of the Serpens region (see caption of Fig.~\ref{serpens}).
The dashed arrows indicate the measured (absolute) proper motions.
The dotted triangles drawn around the proper motion vectors represent the
uncertainty in the magnitude and direction of the proper motions. A dotted 
arrow is used to represent the proper motion of the feature with label 
number ''3'' to indicate that has the most uncertainty, with SNR $\leq$ 1.  
The magnitude scale of the proper motions is given at the bottom left-hand
corner of the panel. The plot axes give absolute R.A. and DEC coordinates.}
\label{rno15}
\end{figure*}

\subsection{RNO~15-FIR}

The upper panel of Fig.~\ref{rno15} presents a H$_{2}$ 2.12~$\mu$m image of the
RNO~15-FIR outflow region \citep{Dav97a} with contours of the 
CO $3~\to~2$ integrated intensity map by \citet{Dav97b} overlaid. 
The accuracy of the positional registration of the H$_{2}$ and CO images is 
a few arcseconds. Several bright, compact H$_{2}$ knots 
lie along the axis of the CO outflow, and are found to have good spatial
correspondence with the peaks of the CO emission. These observations indicate
the presence of a highly collimated (opening angle $\leq$ 10\degr) jet/outflow, 
directed at P.A.$\approx$32\degr . The IRAS position of RNO~15-FIR lies 
close to the CO outflow center. The bolometric luminosity of the region
is $\leq 10$~L$_{\odot}$, and both the measured envelope mass (0.9~M$_{\odot}$)
and evolution models indicate a protostellar mass $\leq$ 1~M$_{\odot}$
\citep{Fro03, Fro05}.

The red cross denotes the absolute position of the 22.2~GHz water masers
(accurate to 2~mas\footnote{The uncertainty of the absolute position is dominated 
by the phase-reference calibrator position uncertainty, as the position of the 
reference maser spot in the phase-referenced map is derived with an accuracy of 
a few tenths of mas.}) derived by means of our phase-reference VLBA
observations. The VLBA water maser map is shown in the lower panel of 
Fig.~\ref{rno15}. We detected four maser
features and three of them were persistent over the two VLBA epochs.
Taking advantage of the phase referencing technique used in 
the observations of this source, for the three persistent maser features, we
could derive the {\it absolute} proper motions. The proper motions
shown in Fig.~\ref{rno15} have been corrected for the earth and solar motions,
and for the galactic rotation, which together account for 
$\approx$ 40 -- 50\% of the apparent proper motion of the reference spot.
A simple flat circular rotation model of the Galaxy was assumed, using
a galactic center distance of 8.5~kpc and a rotation velocity at the solar 
circle of 220~km~s$^{-1}$. The distance assumed for RNO~15-FIR is 0.35~kpc.

The four maser features are distributed 
along a line that coincides with the direction of their motion. The 
most widely separated features are only 1.4~AU apart. 
The magnitude of the measured proper motions is in the range of 
10 -- 40~km~s$^{-1}$. The line of sight velocities of the maser
features are within 1 -- 2~km~s$^{-1}$ from the LSR velocity of the
ambient gas, $\approx$4.7~km~s$^{-1}$.

\section{Discussion}   

\subsection{Serpens SMM1}

The low-mass YSO SMM1 in the Serpens molecular core is the source for 
which our VLBA observations produced the highest number of detected maser
features and measured proper motions.
SMM1 is the brightest of the mm sources detected in the
Serpens protocluster by \citet{Tes98}.
OVRO continuum mm observations by \citet{Hog99} resolve
the dust emission from the envelope surrounding the central protostar: the
best fit to the data is obtained with an extended envelope, having a radius
of 8000~AU and a mass of 
8.7~M$_{\odot}$, plus the contribution of a compact (100~AU radius)
source with a mass of 0.9~M$_{\odot}$. Since the calculated envelope mass 
exceeds the upper limit on the protostellar mass of 3.9~M$_{\odot}$ (derived 
from the bolometric luminosity), \citet{Hog99} conclude that SMM1 is likely 
a Class 0 protostar, with most of the final mass to be still accreted.
\citet{Dav99} observed the submillimeter continuum emission of the Serpens
region using the SCUBA bolometer array camera mounted on the JCMT. Fitting the
continuum emission of SMM1 from the mid-infrared to the millimetre wavelengths
with an optically thin blackbody curve, these authors obtained a value for the 
best fit temperature of 38~K. Since this temperature is significantly higher
than the typical value ($\approx$20~K) observed towards Class~0 YSOs, 
\citet{Dav99} suggest that SMM1 may be found at an intermediate evolutionary 
stage between Classes 0 and I. 

The Class~0 and/or Class~I evolutionary stages are the ones where most of 
the protostellar mass is accreted, and YSOs belonging to these classes should 
be surrounded by a torus- or disk-like distribution of matter, accreting onto 
the protostellar surface.  For SMM1, \citet{Bro00} have resolved the submillimeter 
dust emission surrounding the YSO, using the James~Clerk~Maxwell~Telescope 
-- Caltech~Submillimeter~Observatory \ single-baseline interferometer on Mauna 
Kea (Hawaii).  By fitting a power-law disk model to the data, these
authors derive a lower limit to the disk mass of  \ 0.1~M$_{\odot}$
and a disk radius of $\approx$90~AU. 

The position angle on the sky of the accretion disk or torus is 
expected to be perpendicular to the axis of the radio jet observed 
towards SMM1 (see Fig.~\ref{serpens}d), whose spatial structure has 
been imaged using the VLA with a linear resolution ($\approx$60~AU),
comparable with the estimated disk radius.  The two observed clusters 
of water maser features are oriented on the sky along a direction 
approximately perpendicular to the axis of the radio jet 
(Figs.\ref{serpens}d~and~\ref{serpens}e.  The location of the water 
masers (at the origin of the radio jet), the diameter of the maser 
distribution ($\approx$25~AU, smaller than the expected disk size) 
and its orientation on the sky, are all compatible with the scenario 
where the maser emission originates close to the inner regions of 
the accretion disk. 

The physical conditions for 22.2~GHz water maser excitation require high
gas density (n$_{H_{2}} \sim 10^{9}$~cm$^{-3}$) and temperature 
($T_{kin} \geq 400~K$). Excitation models predict that these conditions can be
reached in the shocked layers of gas behind both high-velocity 
($\geq$ 50~km~s$^{-1}$) dissociative (J-type) \citep{Eli89} and slow 
($\leq$ 50~km~s$^{-1}$) non-dissociative (C-type) \citep{Kau96} shocks, 
propagating in dense regions (pre-shock density is $\geq 10^{7}$~cm$^{-3}$).  
If the water maser emission towards SMM1 originates from gas in the
accretion disk at distances of (only) 10--20~AU from the protostar, one might
argue that the temperature and the density of the environment would be
sufficiently high to excite the water masers without the need of shock heating
and compression. In this view the water masers might trace the 
kinematics of the gas in the disk. However, as we discuss in the following,
the detected maser features appear to
move {\em faster} than expected if their acceleration were due to the
gravitational field of the low-mass YSO in SMM1.

As already mentioned, all the maser features have line of sight velocities 
red-shifted by more than \ 10~km~s$^{-1}$ with respect to 
the LSR velocity of the ambient gas ($\approx$8.5~km~s$^{-1}$), and 
the measured relative transverse velocities have magnitudes in the range
\ 30 -- 40~km~s$^{-1}$.
The distance to SMM1 is not accurately known.
Adopting the maximum value reported in the literature of 400~pc \citep{Hur96},
an upper limit for the bolometric luminosity of \ 77~L$_{\odot}$ is
derived, which translates into an upper limit for the protostellar mass (taking
the mass of a main-sequence star with the same luminosity) of  \ 3.9~M$_{\odot}$ 
\citep{Hog99}.  This value is probably larger than the likely stellar 
mass, as SMM1 is a YSO in an early stage of evolution and it is thus expected 
to be overluminous compared to the final main-sequence luminosity.  The velocity 
of gas rotating in a Keplerian disk around a \ 4~M$_{\odot}$ object at a distance 
of \ 12.5~AU (the radius of the observed water maser distribution) is 
\ $<$ 20~km~s$^{-1}$. Comparing this value with the measured transverse 
velocities, one sees that, even assuming an implausibly high value for the 
mass of the YSO, the expected velocities for Keplerian rotation are lower than 
those measured.
  
If the masers originate from the gas in a disk or toroid, then they likely 
trace a velocity field other than Keplerian rotation. The fact that the line 
of sight velocities of the maser features are strongly red-shifted, is a hint for 
identifying the driver of the maser motion. Fig.~\ref{serpens}d shows that 
the emission from the northwestern lobe of the radio jet observed in Serpens 
SMM1 is stronger and has a better defined structure than the emission associated 
with the southeastern lobe. OVRO molecular line observations (angular 
resolution of \ 1$^{''}$ -- 5$^{''}$) by \citet{Hog99} towards Serpens SMM1 
support the presence of a bipolar molecular outflow oriented along the same 
direction of the radio jet, with the highest velocity and more collimated flows 
being red-shifted towards the northwest and blue-shifted towards the southeast. 
Although the angular resolution of the OVRO molecular line data is worse (by a 
factor 3 -- 10) than that of the 3.6~cm VLA-A observations of the radio jet, the
angular scales sampled with the two different outflow tracers are sufficiently 
similar to infer the inclination angle of the jet axis with the line of
sight: the ionized gas associated with the {\em northwestern} jet lobe 
corresponds to the red-shifted lobe of the molecular outflow and, hence, 
has to be moving {\em away} from us. 

\begin{figure}
\centering
\includegraphics[width=9.5cm]{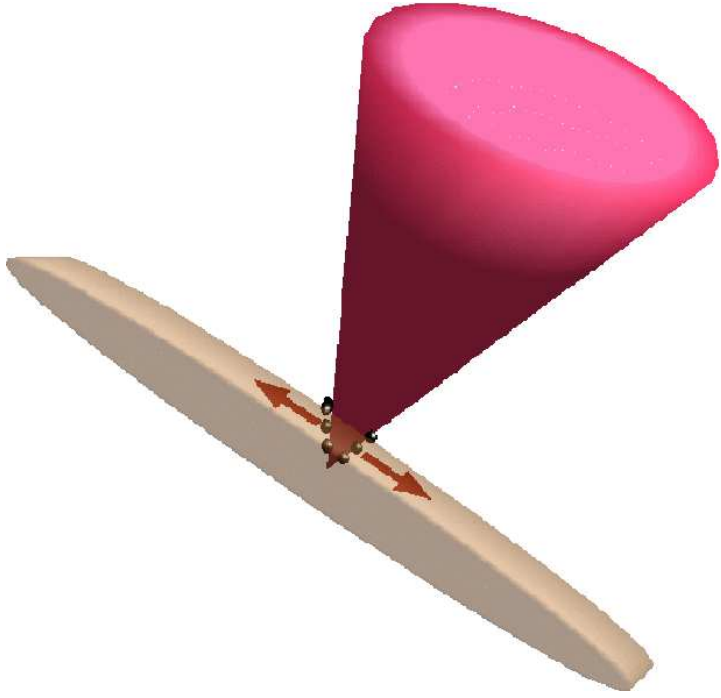}
\caption{Sketch of the red-shifted jet lobe impacting with the disk/torus material
close to the
low-mass YSO and exciting the water masers. The disk/torus is transparent
at cm wavelengths. The small black spheres along
the sides of the jet
lobe indicate the expected location of the water masers. The red arrows indicate
the measured proper motions of the maser spots transverse to the jet axis.}
\label{mjd_sketch}
\end{figure}

The red-shifted velocities of the water masers might be naturally explained if 
they originated from dense shocked material at the interface between the
red outflow lobe and the circumstellar material.  Our proposed interpretation 
is that the water maser emission emerges at the very base of the jet, where one 
can imagine that the jet gas interacts with the material surrounding the YSO in 
a toroid or ``thick'' disk. It is worth noting that, with the VLA angular 
resolution of $\approx$60~AU, the central component of the radio jet (see 
Fig.~\ref{serpens}d) is unresolved along its minor axis. At the base of the jet
where the interaction with the disk may occur, the (sky-projected) transverse  
size of the jet could be even smaller than the separation ($\approx$25~AU) 
between the two observed clusters of water maser features, which might trace 
the shocked molecular gas along the walls of the cavity excavated by the outflow 
in upper layers of the toroid.  
One might expect the water masers to move significantly faster 
than the gas of the Keplerian disk, and mainly trace the outflow motion 
induced by the jet on the surrounding molecular gas. 
Figure~\ref{mjd_sketch} presents a sketch of the 
proposed interpretation.

This interpretation does not conflict with the observational result that 
the measured maser proper motions are transverse to the jet axis, since, 
measuring only {\em relative} transverse velocities prevents us from 
detecting any common motion of all the maser features (eventually 
directed along the jet axis). To account for the detected transverse 
motion of the two maser clusters (with magnitude of 30--40~km~s$^{-1}$), 
a lateral expanding motion of the jet of amplitude \ $>$ 15--20~km~s$^{-1}$ \ 
is required. Since a collimated outflow should have the parallel (to the
jet axis) component of velocity significantly larger than the transverse
component, we speculate that the maser spot motion along the jet axis 
should be \ $\gg$ 20~km~s$^{-1}$.  Then, the (relatively small) redshifts 
of \ 10--15~km~s$^{-1}$ measured for the water masers might be explained 
if the axis of the jet were close to the plane of the sky. This proposed scenario 
is partially supported by the high velocities found for the ionized gas. 
By means of multi-epoch 3.6~cm VLA-A observations, \citet{Cur93} have 
measured the proper motion of emission knots in the jet lobe deriving 
velocities of $\approx$200~km~s$^{-1}$.

It is clear that this interpretation can and should be tested by
measurements of {\it absolute} proper motions of the maser features.
Further VLBI observations will also allow us to establish if the 
spatial and velocity distribution we have now observed continues over
time, as one would expect in the case that, as we are suggesting, 
the water masers trace a stable kinematic structure. 

\subsection{RNO~15-FIR}

Our results towards RNO~15-FIR are a good example of the advantages of 
performing VLBI observations in phase-reference mode. Although only 
four maser features were detected, the measurement of {\em absolute} 
proper motions for three of them offers a rather straightforward 
interpretation of the water maser kinematics.  Fig.~\ref{rno15} shows 
that the absolute position of the maser emission is close to
the centre of the molecular outflow observed in CO on angular scales 
much larger than those traced by the water masers. The absolute maser
velocities are all found to be aligned along a direction that agrees 
well with the axis of the CO molecular outflow and the line connecting 
the bright, compact H$_{2}$ knots. Our suggested interpretation is that 
the water masers are tracing the innermost portion of the molecular 
outflow, likely emerging from condensations of material shocked
by the passage of jet material from the low-mass YSO. 

VLA observations by \citet{Mee98} towards RNO~15-FIR did not detect 
any radio continuum emission at 4.8 and 8.4~GHz (with upper limits, 
of 0.12 and 0.24~mJy~beam$^{-1}$, respectively).  If we compare this 
result with the VLA observations towards Serpens SMM1, where an ionized 
jet is detected with peak fluxes of a few mJy, we would argue that in 
RNO~15-FIR the volume of the circumstellar gas ionized by interaction 
with the jet is too small to produce detectable free-free emission. 
The values of mass, momentum and energy derived for the CO molecular 
outflow in RNO~15-FIR by \citet{Dav97b} are significantly lower than 
those of other outflows from heavily embedded, low-mass YSOs. 

Using ISOPHOT and SCUBA measurements of this region, \citet{Fro03} 
determined the spectral energy distribution (SED) from submillimeter 
to far-infrared wavelengths, obtaining a ratio of \ L$_{smm}$/L$_{bol}$ \ 
which classifies RNO~15-FIR as a Class~0 source. However, the relatively 
high value of bolometric temperature, 44.6~K, resulting from the fit 
to the SED, may suggest that RNO~15-FIR is in a later, Class~0/Class~I 
transition, evolutionary stage. Based on the evolutionary model of 
\citet{Fro03}, which combines the unification scheme \citep{Smi00} 
with the framework for protostellar envelopes \citep{Mye98}, 
RNO~15-FIR and Serpens SMM1 should have approximately the same age, 
$\approx$3 -- 3.5 $\times$ 10$^{4}$~yr. The quite different bolometric 
luminosities of the two sources can be explained in terms of different 
values for the mass of the two YSOs. RNO~15-FIR is believed to be less 
massive than Serpens SMM1 by a factor of a few. The corresponding
lower values of mass accretion rate expected for RNO~15-FIR would translate
into lower mass ejection rates, which might partly explain the strikingly 
different properties of the radio continuum and molecular line emissions 
observed for the outflows associated with the two sources.

In contrast to what we observed in Serpens SMM1, where the water maser 
spatial distribution is transverse to the axis of the radio jet, the 
few water maser features detected in RNO~15-FIR are aligned along a 
direction parallel to the orientation of the molecular outflow. Based 
on this, we propose that the maser emission originates from dense clumps 
of gas displaced along the axis of the jet emitted by the YSO. Over 
comparable protostellar lifetimes, whereas the relatively powerful jet 
emitted by Serpens SMM1 succeeded in sweeping away from its course the
densest portion (n$_{H_{2}}$ $\geq 10^{7}$~cm$^{-3}$) of the circumstellar 
material and creating a tunnel of lower density (n$_{H}$ $\sim 10^{4}$ -- 
$10^{5}$~cm$^{-3}$) ionized gas, that is not the case for the weaker jet 
emerging from RNO~15-FIR, which is undetectable because still confined 
within a very dense envelope of molecular gas. In conclusion, we suggest 
that in RNO~15-FIR the alignment of the water maser features parallel 
to the axis of the molecular outflow indicates that very dense clumps 
of circumstellar material are still to be found on the way along the jet.

\begin{figure}
\centering
\includegraphics[width=14cm]{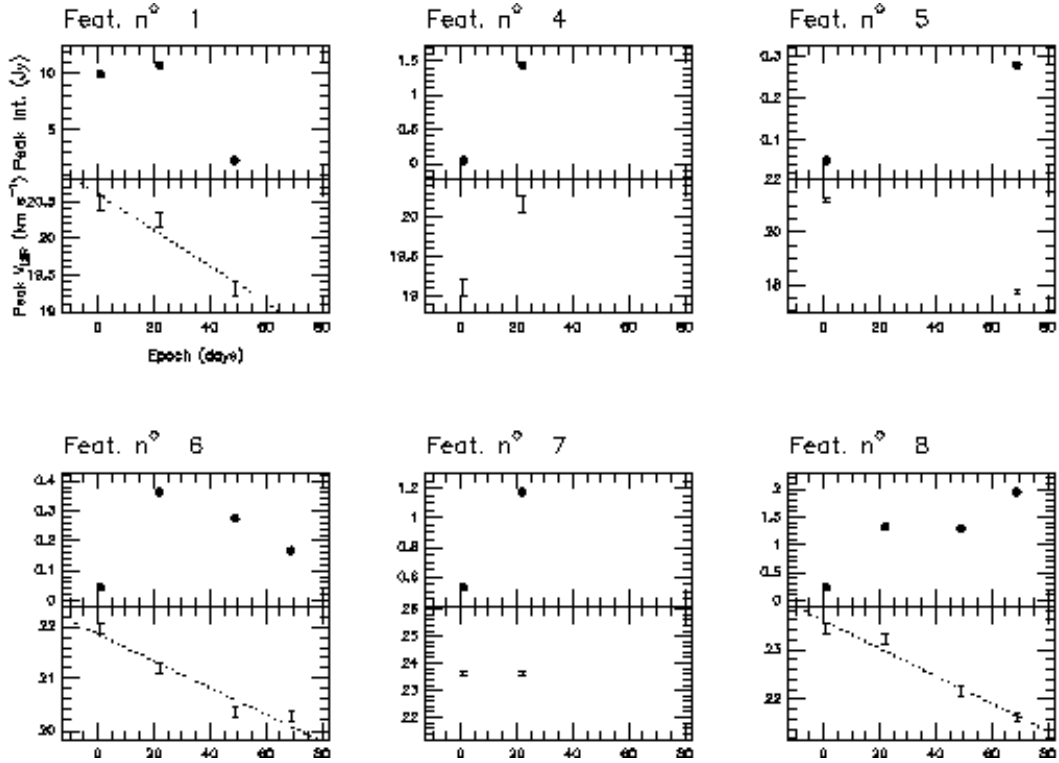}
\caption{Intensity and V$_{LSR}$ maser variability in Serpens SMM1. Each plot
refers to a different maser feature (whose label number is indicated on  
top of the plot) and presents the variation of the peak channel V$_{LSR}$ 
(lower panel) and intensity (upper panel) over time.  The error 
of the line of sight velocity is taken equal to the channel separation of
0.1~km~s$^{-1}$. For features observed for more than three epochs, the dashed
line gives the (least squares) linear fit of the variation of peak channel 
V$_{LSR}$ with time.}
\label{sp_var_smm1}
\end{figure}
 
\subsection{Maser feature structure and variability}

Figures~\ref{sp_epo_smm1}~and~\ref{sp_epo_rno15} show the maser features' 
structure in Serpens SMM1 and RNO~15-FIR respectively. It is interesting 
to note that in both sources all the spatially extended features present 
spot distributions with similar orientation, which is approximately parallel 
to the axis of the radio (for SMM1) or molecular (for RNO~15-FIR) jet 
observed at larger scales (see Figures~\ref{serpens}~and~\ref{rno15}). 
The previous discussion has suggested that in both sources the maser emission 
is excited by the interaction of the jet with dense surrounding gas. If the 
water maser features originate behind shocks, a gradient of density, 
temperature and velocity is naturally expected {\em along} the shock 
propagation direction, which, for jet induced shocks, is parallel to the
jet motion. The fact that the maser features are observed to be spatially
extended with orientation close to the jet axis therefore reinforces our
interpretation that the origin and kinematics of the water masers stem from
the YSO jet.

Figure~\ref{sp_epo_smm1} shows that the features in Serpens SMM1 that 
persist over three or more epochs (the ones with label numbers ``1'', ``6'' 
and ``8'') have a stably decreasing V$_{LSR}$ (i.e. spot and feature symbols 
have colours changing over time from red to blue).  Figure~\ref{sp_var_smm1} 
shows that the velocity of the emission peak of these features decreases by 
\ 1.5 -- 2~km~s$^{-1}$ over a time span of \ 50 -- 70~days, corresponding 
to a deceleration of $\approx$0.03~cm~s$^{-2}$. This value is too high to 
result from the gravitational forces of the YSO in SMM1.  Assuming a 
(maximum) central mass of \ 4~M$_{\odot}$ and a (minimum) distance 
of 10~AU, we estimate an upper limit to the gravitational acceleration of
0.006~cm~s$^{-2}$. In agreement with the proposed ''jet'' origin for the maser
emission in SMM1, a possible explanation for the measured features' 
decelerations might be that the shocks responsible for the maser excitation
are slowed down as they proceed through the dense material surrounding the
YSO. The law of variation of the velocity with time depends on the 
nature of the shock (i.e., its cooling efficiency)
and/or whether viscosity or turbulence contribute to the braking. The
available data are insufficient to discriminate among the different 
possibilities.

As already noted in Section~4.2, Figure~\ref{prop-mot} shows that, in the 
case of feature ''6'', the variation of sky-projected positions with
time is fitted better by a second-order
polynomial (i.e. a uniformly accelerated motion) than by a linear fit (i.e. a 
constant motion). The RMS of the residuals (both for the motion along the
R.A. and Dec axis) is only 1--2~$\mu$as for the quadratic fit, 
whereas for the linear fit it is of the order of tens of microarcseconds.
Using $V_{0}$ and $A_{0}$ respectively to indicate the initial (at first-epoch) 
velocity and the constant acceleration of the uniformly accelerated motion, 
the values derived with the quadratic fit are: 
$V_{0}$ = 42~km~s$^{-1}$,
$A_{0}$ = -15~km~s$^{-1}$~month$^{-1}$, for the motion along the R.A. axis;    
$V_{0}$ = 13~km~s$^{-1}$,
$A_{0}$ = 11~km~s$^{-1}$~month$^{-1}$, for the motion along the Dec axis.    
The ''average'' velocity of the uniformly accelerated motion 
(25~km~s$^{-1}$ and 26~km~s$^{-1}$ along, respectively, 
the R.A. and Dec axis) is consistent with the proper motion value
derived with the linear fit (see Table~\ref{features}).

The motion of feature ''6'' is clearly decelerating eastward and 
accelerating northward. Along both directions, the magnitude of the 
derived acceleration is high enough to let the initial (first-epoch) 
and the final (last-epoch) velocity differ by a factor of $\approx$2 
with respect to the average (central) velocity.  The deceleration measured 
for feature ''6'' along the line of sight is $\approx$0.8~km~s$^{-1}$~month$^{-1}$, 
more than one order of magnitude smaller than the magnitude of the 
accelerations derived for the motions along the R.A. and the Dec axis. 
This is in qualitative agreement with the jet model proposed to explain 
the maser kinematics in Serpens SMM1, which requires the jet to be nearly in
the plane of the sky. We then expect that the maser features' accelerations 
(decelerations) are also close to the plane of the sky. In interpreting the 
sky-projected motion of feature ''6'' one has to keep in mind that it
is measured {\em relative} to another feature (label number ''8'').
The derived acceleration of feature ''6'' is actually the difference 
of the accelerations (decelerations) of features ''6'' and ''8'', both along 
the R.A. and Dec axis. Our interpretation is that, since the two features
are moving across different portions of the dense material surrounding the 
YSO, they are {\em decelerating} at different rates, the eastward (northward) 
motion of feature ''6'' changing more (less) rapidly than for feature ''8''.   

\section{Conclusions}

This article reports the first results of a long-term project whose aim,
by using the VLBA, is to survey the 22.2~GHz water maser emission in a
selected sample of low-mass YSOs. The sample includes objects in different
evolutionary stages (Class 0, Class 0/Class I transition, Class I) so that
the water maser spatial and velocity structure may be correlated with the
YSO age. We present VLBA water maser data from 2003 from a cluster of 
low-mass YSOs in the Serpens molecular core and from the low-mass YSO 
RNO~15-FIR.  

Towards Serpens SMM1, the most intense submillimeter source of the
Serpens molecular core, the maser emission is found to originate from 
two clusters of 
strongly red-shifted (more than \ 10~km~s$^{-1}$) features 
separated by $\approx$25~AU. The measured relative transverse velocities of the
maser features are parallel to the cluster-connecting line, which in turn
is approximately perpendicular to the axis of the radio jet observed towards SMM1
on length scales of hundreds of AU. Based on their spatial distribution, water
masers might originate in the accretion disk surrounding the YSO. However, the measured 
maser (line of sight and transverse) velocities appear too large to be compatible 
with Keplerian rotation around a central mass $\leq$ 4~M$_{\odot}$. We suggest 
that the water maser emission originates at the very base of the radio jet,
tracing the interaction region of the red-shifted lobe of the jet with the dense 
material of the accretion disk. Since the jet is much faster than the gas rotating
in the disk, the water maser kinematics are driven by the jet.
    The line of sight velocities of
    several features decrease at a rate of $\approx$1~km~s$^{-1}$~
    month$^{-1}$ and for one feature (label number ''6'') the sky-projected
    motion (relative to another feature) appears to be accelerated 
    (decelerated)
    at a rate of $\approx$10--15~km~s$^{-1}$~month$^{-1}$. The 
    proposed interpretation 
    is that the shocks, excited by the propagating jet and harboring 
    the maser emission, are slowed down as 
    they proceed through the dense material surrounding the YSO.

Only a few (four) maser features were detected towards RNO~15-FIR. However, since
{\em absolute} proper motions were measured, the available data is sufficient to 
suggest the position of the water maser birthplace. The water maser features are 
distributed along a line, whose orientation on the sky coincides with the
common direction of
 all the measured absolute velocities. Such a spatial and
velocity distribution is expected if water masers trace a tightly collimated flow.
Since the derived VLBA water maser absolute position locates them
just at the center of the bipolar molecular outflow observed towards RNO~15-FIR (on
angular scales of tens of arcsecs), and since the axis of the molecular outflow is parallel with the direction of elongation and motion of the maser features, we suggest
that the water masers trace the innermost portion of the molecular outflow.

On the basis of the measured bolometric luminosities and temperatures, evolutionary
models indicate that RNO~15-FIR and Serpens SMM1 share a comparable evolutionary 
stage, both of them still being Class 0 sources, even if more evolved than an average 
Class 0 protostar. Despite this, we observe significant differences in the water
maser spatial and velocity distribution between the two sources. One possible explanation 
may be related to the different value of protostellar mass
of the two YSOs. SMM1 is likely more massive than RNO~15-FIR by a factor of 
at least a few. A higher value of protostellar mass could imply correspondingly
higher values of mass accretion (and ejection) rates. If the jet emitted by the
YSO is sufficiently powerful to sweep away the densest portions of 
circumstellar gas, eventually the molecular gas density along the jet axis 
becomes too low (n$_{H_{2}}$ $< 10^{7}$~cm$^{-3}$) for the excitation
of the water masers. In that case, however, water maser emission can still 
occur if the jet interacts with the dense material of the accretion disk.
At present, all this is highly speculative.  To put these arguments on firm ground 
it will be necessary to expand our observations to larger samples of objects of 
various masses and in various evolutionary stages. Additionally, our proposed 
interpretation of the water maser morphology and kinematics in the Serpens SMM1 
source need to be put on firm ground by measuring the {\it absolute} location and 
proper motions of the maser spots.

\begin{acknowledgements}
It is a pleasure to thank Prof. Tetsuo Sasao for allowing us to use his program
for the computation of apparent proper motions due to annual parallax, solar
motion with respect to the LSR, and galactic rotation. We also thank the 
anonymous referee for his/her warning about the evidence for maser feature 
accelerations contained in our data.  
\end{acknowledgements}

\bibliographystyle{aa}
\bibliography{biblio}

\end{document}